\documentclass[11pt,a4paper]{article}
\usepackage{jcappub}
\usepackage{color}
\usepackage{graphicx}
\usepackage{times}
\usepackage{amssymb}
\usepackage{amsmath}
\usepackage{url}

\title{Cosmic Mach Number: A Sensitive Probe for the Growth of
Structure}

\author[a]{Yin-Zhe Ma,}
\author[b,a]{Jeremiah P. Ostriker,}
\author[c,d]{Gong-Bo Zhao}

\affiliation[a]{Kavli Institute for Cosmology and Institute of
Astronomy, University of Cambridge, Madingley Road, Cambridge, CB3
0HA, UK} \affiliation[b]{Department of Astrophysical Sciences,
Princeton University, 4 Ivy Lane, Princeton, NJ, U.S.A}
\affiliation[c]{Institute of Cosmology and Gravitation, University
of Portsmouth, Dennis Sciama Building, Portsmouth, PO1 3FX, UK}
\affiliation[d]{National Astronomy Observatories, Chinese Academy
of Science, A20 Datun Road, Chaoyang District, Beijing, China}

\emailAdd{mayinzhe@phas.ubc.ca}
\emailAdd{ostriker@princeton.edu}
\emailAdd{Gong-Bo.Zhao@port.ac.uk}

\abstract{
We investigate the potential power of the Cosmic Mach Number
(CMN), which is the ratio between the mean velocity and the
velocity dispersion of galaxies as a function of cosmic scales, to
constrain cosmologies. We first measure the CMN from $4$ catalogs
of galaxy peculiar velocity surveys at low redshift
($z\in[0.002,0.03]$), and use them to contrast cosmological
models. Overall, current data is consistent with the WMAP7
$\Lambda$CDM model. We find that the CMN is highly sensitive to
the growth of structure on scales $k\in[0.01,0.1]$ h/Mpc in
Fourier space. Therefore, modified gravity models, and models with
massive neutrinos, in which the structure growth generically
deviates from that of the $\Lambda$CDM model in a scale-dependent
way, can be well differentiated from the $\Lambda$CDM model by
using future CMN data.}

\arxivnumber{1106.3327}


\begin{document}
\maketitle

\section{Introduction}

The Cosmic Mach Number (hereafter CMN) can provide a robust
measure of the shape and growth rate of the peculiar velocity
power spectrum of the galaxies in the universe. One considers a
region of a given size $r$ in the universe, and compares the bulk
motion of the sphere with the random velocities within that
region. The bulk motion provides a measurement of the forces on
the region from irregularities external to it, so it measures the
amplitude of perturbations on scales much larger than the region,
whereas the random motions within the comoving region reflect
gravitational perturbations on scales smaller than $r$. Thus their
ratio, $M(r)$, depends on the shape of the perturbation spectrum
while independent of its amplitude.

The concept was introduced by \citet{Ostriker90} as a
cosmological metric that would be relatively independent of the
``bias'' of the test particles being observed and also relatively
independent of the then quite uncertain perturbation amplitude.
They concluded that although the existing data were poor, they
gave estimates of the CMN that appeared to be inconsistent with
the then popular CDM model with $\Omega_{m}=1$ but seemed to
prefer the open universe model instead. In a certain sense, the
application of this test, correctly predicted the currently best
validated cosmological models with a value of $\Omega_{m}$ in the
range of $0.2-0.3$.

Subsequent to the original paper, \citet{Strauss93} found again
that the CDM models with $\Omega_{m} = 1$ remained inconsistent
with the better data they used, but some non-standard models
passed the test (see also \citet{Suto90,Suto92}).
\citet{Nagamine01} looked at $\Lambda$CDM models and found
better agreement, but the then $\Lambda$CDM model with
$\Omega_{m}=0.37$ again produced too high values of $M$ over the
range of $3-40 \mathrm{Mpc/h}$, whereas a model with $\Omega_{m}
\sim 0.2$ (actually closer to WMAP 7-yr constraint
(\citet{Komatsu11})) was more consistent with the observations.
In addition, there are various other papers discussing the issues
related to the CMN, such as non-linear clustering properties of
dark matter on the CMN measurement (\citet{Gramann95}), and
constraints on the CMN from Sunyaev-Zeldovich effect
(\citet{Atrio-Barandela04}).

This history leads us to re-examine the issue in light of the much
better knowledge now available from both the new peculiar velocity
data and the range of models remaining plausible given the current
cosmological constraints. In this paper, we will develop a new
statistical tool to measure the CMN from the peculiar velocity
surveys, and investigate the power of the CMN to distinguish
various cosmological models, especially in the aspects of
differentiating the $\Lambda$CDM model from variance with
non-trivial growth function provided by Modified Gravity
(hereafter MG) models, and from models with massive neutrinos.

\begin{figure*}
\centerline{
\includegraphics[bb=0 0 455 315,width=2.5in,keepaspectratio=false]{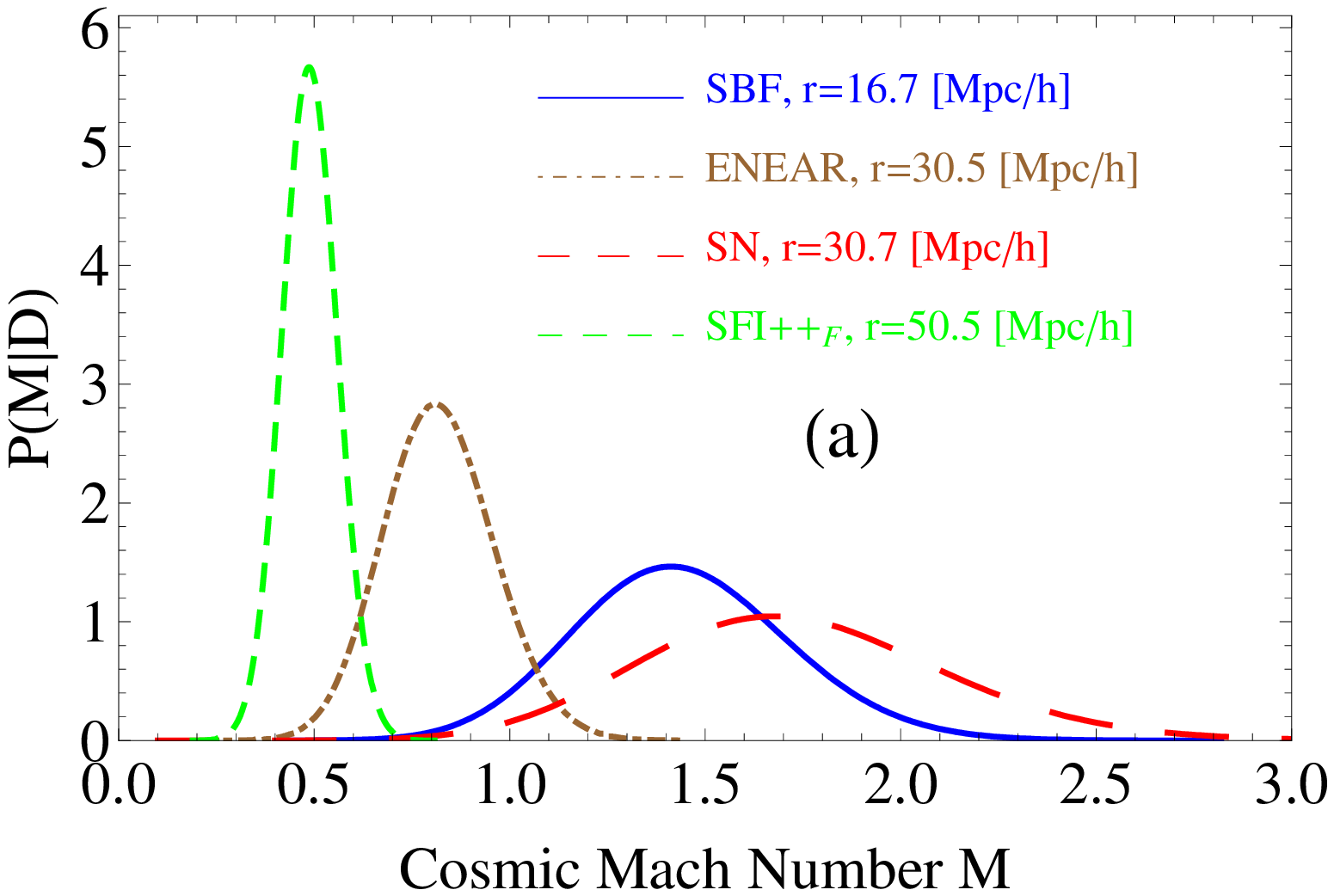}
\includegraphics[bb=0 0 605 405,width=2.4in]{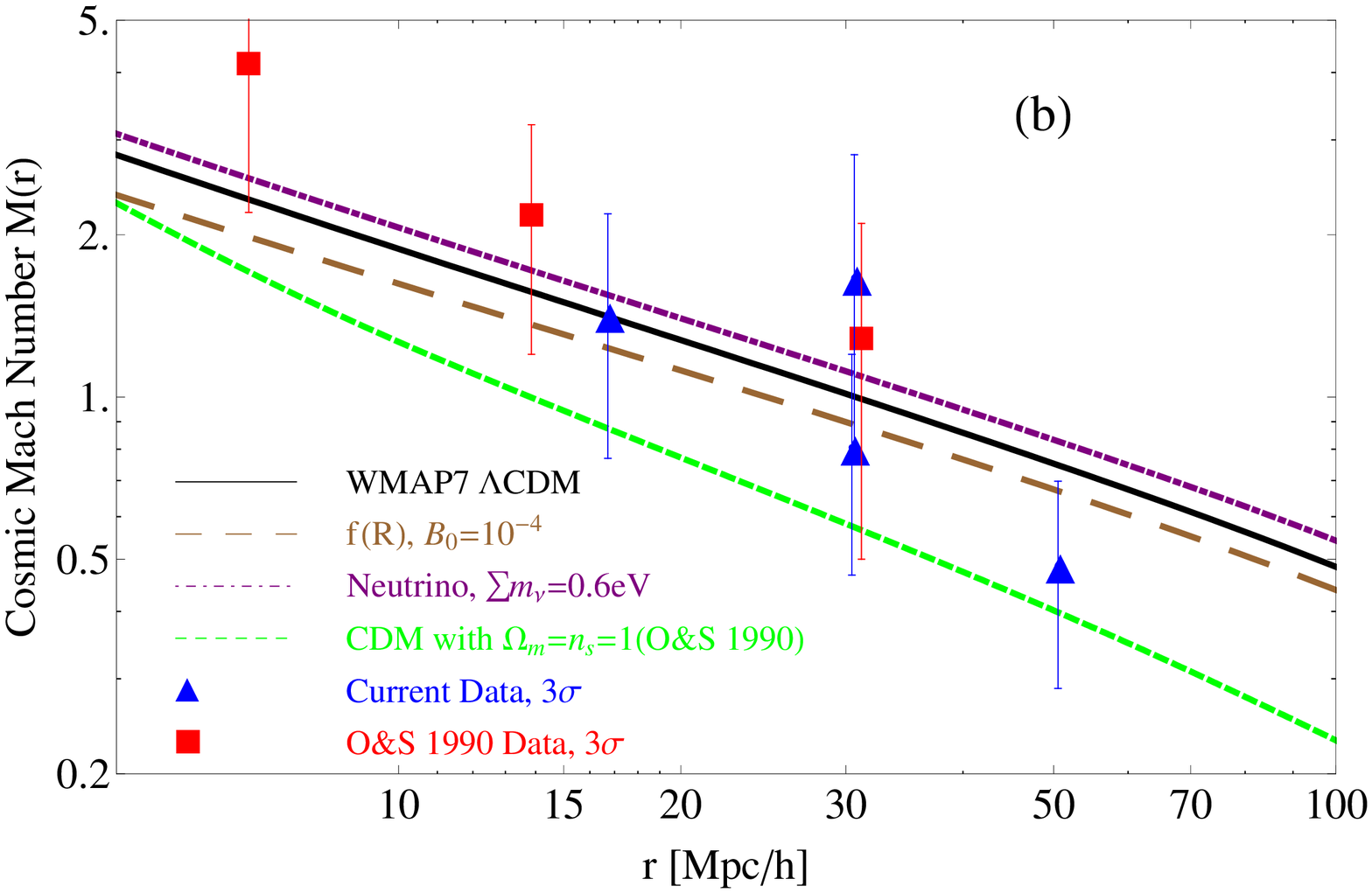}
}
\caption{(a): The 1-D posterior distribution of the CMN $M(r)$
from 4 different catalogs. (b): The comparison between the CMN
data ($3\sigma$ CL.) and theoretic prediction: blue data points
are the CMN data from posterior distributions shown in panel (a);
red data points are the data used in \citet{Ostriker90}. The
black line is the CMN prediction from WMAP 7-yr best-fit values
(\citet{Komatsu11}); the green dashed line is calculated  by
using (1990) `popular' CDM cosmological parameters
$n_{s}=\Omega_{m}=1$ and $h=0.5$ (\citet{Ostriker90}); the brown
dashed and purple dot-dashed lines are the CMN from $f(R)$ model
with $B_{0}=10^{-4}$ and from $\Lambda$CDM model with neutrino
mass $\sum m_{\nu}=0.6$ eV.}  \label{datacompare}
\end{figure*}

\section{Statistics of Cosmic Mach Number} In linear
perturbation theory, the power spectrum of the velocity divergence
($\theta \equiv \nabla \cdot \mathbf{v}$) is related to the power
spectrum of density fluctuations via $P_{\theta
\theta}(k,z)=f^2(k,z)P(k,z)$, where $f(k,z)\equiv-d \ln\delta/d
\ln{(1+z)}$, and $\delta$ is the density perturbation of matter.
Since the data in our application are at very low-redshift, we
assume that they have the same redshift $z=0$ throughout and drop
the $z$ dependence for brevity. The mean square velocity
dispersion $\left\langle \sigma ^{2}(r)\right\rangle$ and mean
square bulk flow $\left\langle V^{2}(r)\right\rangle$ in a window
of size $r$ can be calculated as \citet{Ostriker90,Nagamine01}
\begin{eqnarray}
\left\langle V^{2}(r)\right\rangle  &=&\frac{H_0^{2}}{2\pi ^{2}}%
\int_{0}^{\infty}dk P_{\theta \theta}(k)W(kr) , \nonumber \\
\left\langle \sigma ^{2}(r)\right\rangle  &=&\frac{H_0^{2}}{2\pi ^{2}}%
\int_{0}^{\infty }dk P_{\theta \theta}(k)\left[ 1-W(kr)\right],
\label{msv}
\end{eqnarray}%
where $W(x)=[3(\sin x-x\cos x)/x^{3}]^2$ is a top-hat window
function. Note that $W(x)\sim1(x\lesssim1)$ and $W$ drops to $0$
quickly when $x>1$.

Thus $W$ effectively changes the integral limits of the above
formula as
\begin{eqnarray}\label{eq:vsigma}
\left\langle V^{2}(r)\right\rangle  &\simeq&\frac{H_0^{2}}{2\pi ^{2}}%
\int_{0}^{1/r}dk P_{\theta \theta}(k) , \nonumber \\
\left\langle \sigma ^{2}(r)\right\rangle  &\simeq&\frac{H_0^{2}}{2\pi ^{2}}%
\int_{1/r}^{\infty }dk P_{\theta \theta}(k)=C-\left\langle
V^{2}(r)\right\rangle. \label{msv1}
\end{eqnarray} where $C=\frac{H_0^{2}}{2\pi ^{2}}\int_{0}^{\infty }dk P_{\theta \theta}(k)$ is
a normalization constant. We can see that the velocity dispersion
$\left\langle \sigma^{2}(r)\right\rangle$ can be well approximated
as a flipped signed large scale bulk flow $\left\langle
V^{2}(r)\right\rangle$
with an offset $C$. This is an important feature since it guarantees that no calculation for the nonlinearities is needed to evaluate the velocity dispersion.

The CMN on different scales of the patches is defined as
\begin{equation}
M(r)\equiv \sqrt{\frac {\left\langle
V^{2}(r)\right\rangle}{\left\langle \sigma ^{2}(r)\right\rangle}}
  =\left[\frac{C}{C-\left\langle
V^{2}(r)\right\rangle}-1\right]^{1/2}.
\label{mach_theo}
\end{equation}%
Thus it basically measures the shape of $P_{\theta \theta}$ by
contrasting $\int dk P_{\theta \theta}(k)$ on large, and small
scales. Using Eqs  (\ref{msv}) and (\ref{mach_theo}), one can
reconstruct $P_{\theta \theta}(k)$ up to an overall constant $A$,
\begin{equation}
P_{\theta\theta}(k=1/r)=A\frac{M(r)r^2}{[1+M^2(r)]^2}\frac{dM(r)}{dr}.
\end{equation}
Directly measuring $P_{\theta\theta}$ from peculiar velocity
surveys is challenging, but the measurement of CMN is much easier,
as we show later. Fortunately, CMN has the same information as
$P_{\theta\theta}$, except for the overall amplitude, which can be
easily determined by probes such as the Cosmic Microwave
Background (CMB).


For a peculiar velocity catalog with $N$ objects, the CMN $M$ can
be written as $M=|\mathbf{u}|/\sigma _{\ast }$, where $\mathbf{u}$
denotes the bulk flow velocity, which is a streaming motion of
galaxies towards some particular direction, and $\sigma_{\ast}$
stands for the small scale velocity dispersion. Unfortunately,
neither $\mathbf{u}$ nor $\sigma_{\ast}$ is a direct observable.
For each galaxy peculiar velocity catalog, what we observe is the
line of sight velocity $S_{n}$ with measurement error $\sigma_{n}$
for the $n$th galaxy. Then one can construct a joint likelihood
function for $\mathbf{u}$ and $M$ by contrasting $S_n$ with the
line-of-sight projection of the bulk flow $\hat{r}_{n,i}u_{i}$.
The uncertainty in $(S_{n}-\hat{r}_{n,i}u_{i})$ is simply
$(\sigma_{n}^{2}+\sigma_{\ast }^{2})^{\frac{1}{2}}$, where $\sigma
_{\ast}=|\mathbf{u}|/M$ is given by the definition of the CMN.
Therefore, the likelihood function takes the form of,

\begin{equation}
L(\mathbf{\mathbf{u},}M)= \prod_{n=1}^{N}\frac{1}{\left[ \sigma
_{n}^{2}+(\left\vert \mathbf{u} \right\vert /M)^{2}\right]
^{\frac{1}{2}}}
\times \exp \left(
-\frac{(S_{n}-\hat{r}_{n,i}u_{i})^{2}}{2(\sigma
_{n}^{2}+(\left\vert \mathbf{u} \right\vert /M)^{2})}\right) .
\label{like1}
\end{equation}

One can then marginalize over the 3D bulk flow vector $\mathbf{u}$
to obtain the distribution of $M$ for each survey.

We use four different catalogs of galaxy peculiar velocity
surveys, namely, SBF (\citet{Tonry01}), ENEAR
(\citet{Costa00,Bernardi02,Wenger03}), Type Ia
Supernovae (SN) (\citet{Tonry03}), 
and SFI++$_{F}$ (\citet{Springob07}), to constrain the CMN by
using Eq. (\ref{like1}). To calculate the characteristic depth
$\overline{r}$ of each catalog, we use error-weighted depth as
$\overline{r}=\sum_{n}w_{n}r_{n}/\sum_{n}w_{n}$, where
$w_{n}=1/(\sigma^2_{n}+\sigma^2_{\ast})$. We marginalize over the
`bulk flow' velocity $\mathbf{u}$ in the 4-D parameter space and
obtain the 1-D posterior distribution of $M$ as shown in panel (a)
of Fig. \ref{datacompare}. In the panel (a), one can see that the
distribution of the CMN is very Gaussian, and the width depends
primarily on the number of data entries in each catalog. In
addition, since each catalog probes the CMN on various depths,
different catalogs form a complimentary set of tests of cosmic
structures.

In panel (b) of Fig. \ref{datacompare}, we put together the old
(1990) and current CMN data with various predictions computed from
theoretic models: The [Blue] triangle data points are the current
CMN values and variances computed from likelihood (\ref{like1}) by
using the four-catalogs, which provide more and deeper data than
those used by \citet{Ostriker90} ([Red] square data).

The $\Lambda$CDM model with WMAP 7-yr best-fit parameters
($\Omega_m=0.271$, $h=0.704$, $n_{s}=0.967$, $\Omega_{b}=0.0455$,
[Black] solid line) is mildly consistent with the current CMN data
at $3\sigma$ CL. In comparison, we overplot the theoretical $M(r)$
([Green] dashed line) by using 1990's `popular' CDM parameters.
One can immediately understand the reason why Refs.
\citet{Ostriker90,Strauss93} claimed that there was a strong
conflict between the data ([Red] points) and popular CDM model
([Green] dashed line). There was in fact no inconsistency between
data and a model with $\Omega \approx 0.25$ and $n_{s}=1$.
However, with the up-to-date data and strongly constrained
$\Lambda$CDM cosmology, one can see that the CMN data is
consistent with the $\Lambda$CDM prediction out to scale around
$50 \text{Mpc/h}$.
Note that the small scale CMN can be contaminated by the
nonlinearity of the growth. For instance, using the fitting
formulae of non-linear $P_{\theta \theta}$ proposed in
\citet{Jennings11}, we find that CMN is suppressed about $21\%$
($18\%$) on scale of $15 \textrm{Mpc/h}$ ($20 \textrm{Mpc/h}$).
Thus nonlinearities must be considered when using the CMN data. On
the other hand, numerical simulation can be used to quantify the
effect of non-linearities.

\begin{figure*}
\centerline{
\includegraphics[width=3.0in]{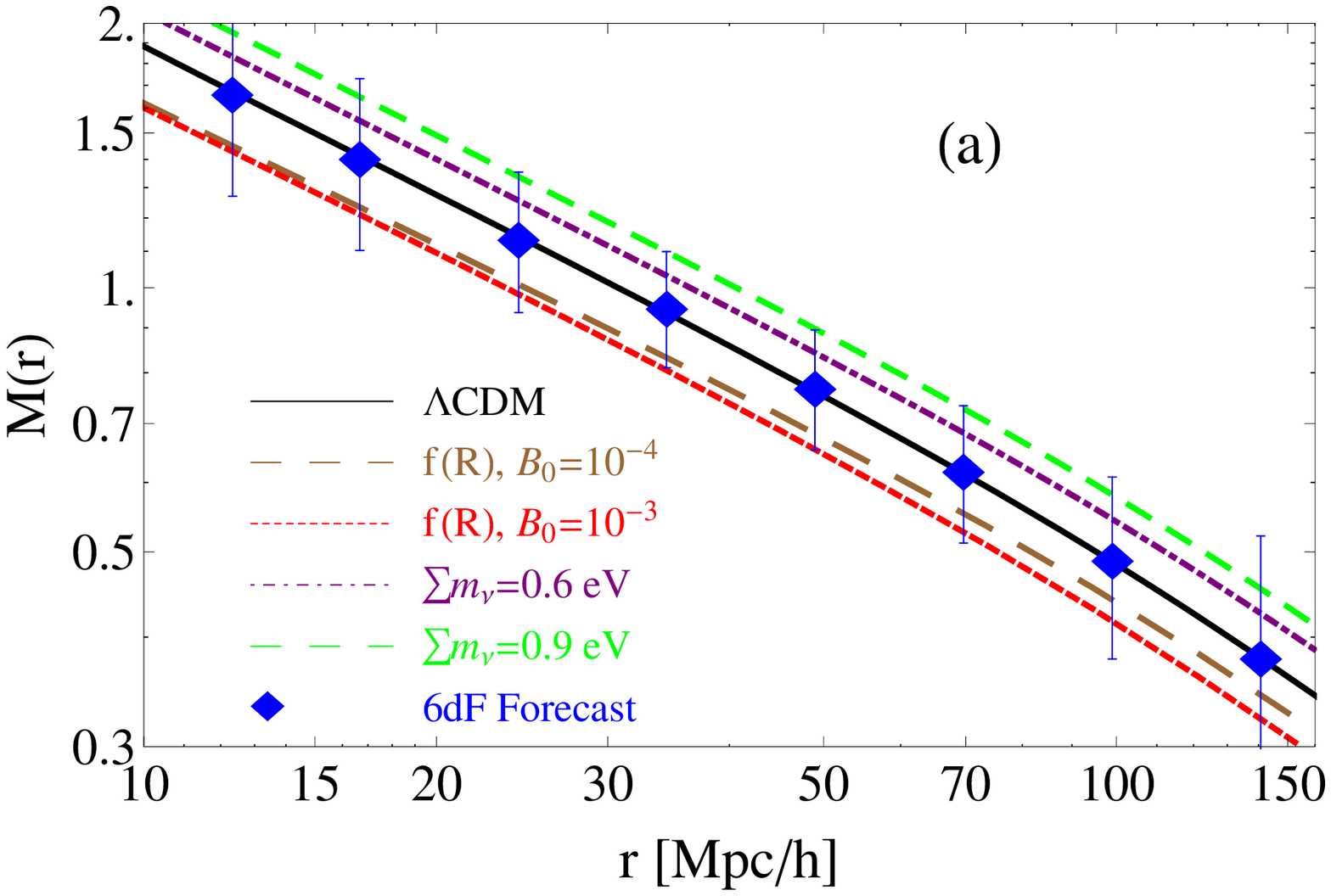}
\includegraphics[width=2.5in]{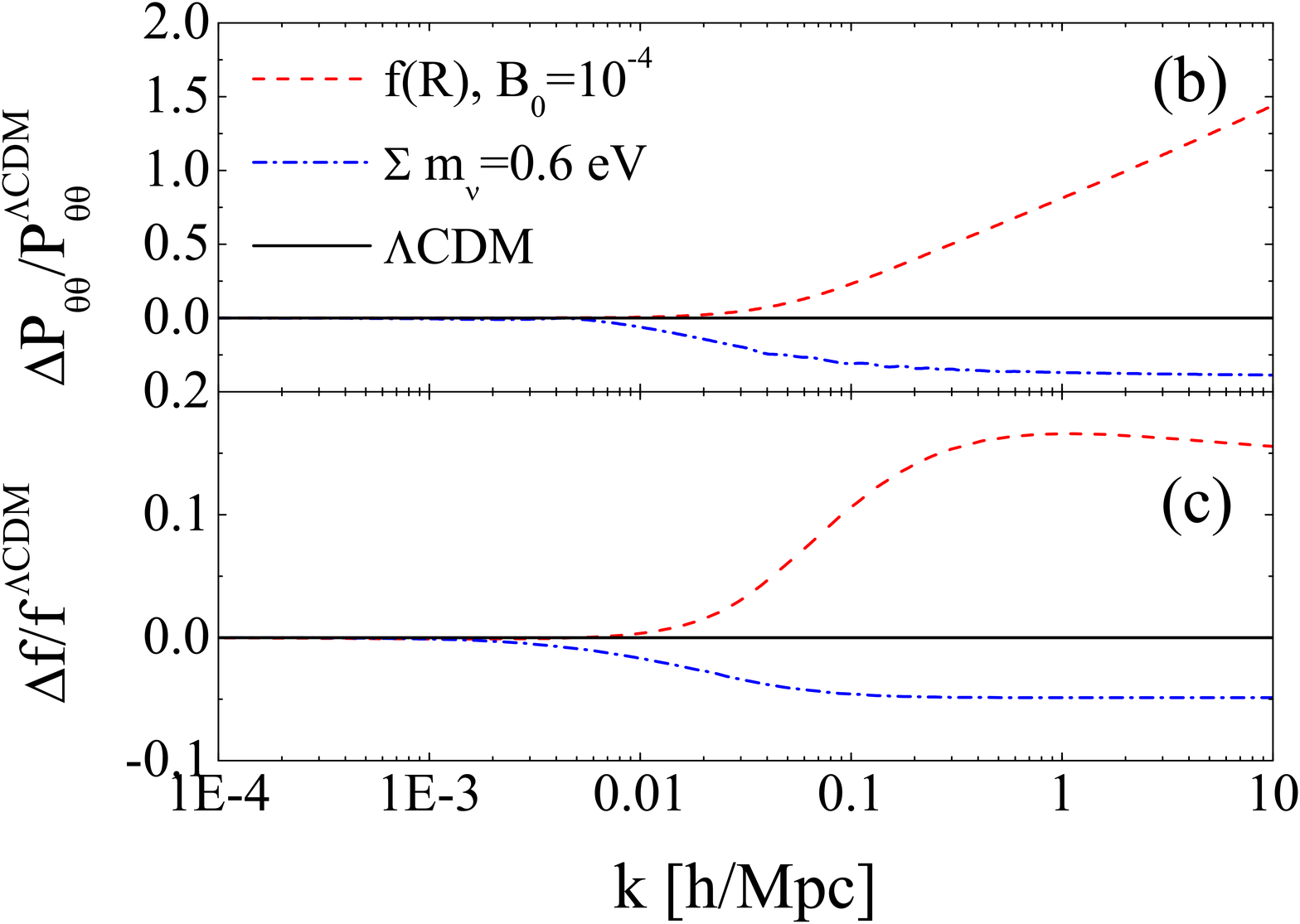}}
\caption{(a): Forecast $1\sigma$ errors of the CMN from 6dF
survey, in which we assume the data is around our local region of
the universe. Beside $\Lambda$CDM model, we plot $B_{0}=10^{-4}$
([Brown] dashed line) and $10^{-3}$ ([Red] dashed line) for the
$f(R)$ model, and $\sum m_{\nu}=0.6 \text{eV}$ ([Purple]
dot-dashed line) and $0.3 \text{eV}$ ([Green] dashed line). (b):
$\Delta P_{\theta \theta}/P_{\theta \theta}$ for $f(R)$ (dashed
line) and massive neutrino (dot-dashed line) models. (c): Same
comparison as (b) but for growth function $\Delta f/f$.}
\label{6dFforecast}
\end{figure*}

To investigate the prospective accuracy of the CMN achievable in
future surveys, we perform a forecast for the on-going 6dF
peculiar velocity survey (\citet{Jones09}). The redshift
distribution of galaxies for the 6dF survey is
$n_{g}(z)=Az^{\gamma }e^{-(z/z_{p})^{\gamma}}$, where $z_{p}\simeq
0.0446 $, $\gamma \simeq 1.6154$ and $A \simeq 622978$
(\citet{Jones09}). It peaks around $z \approx 0.05$ and extends
till $z \approx 0.15$. We assume that the 6dF survey can measure
the peculiar velocity for $12000$ brightest early-type galaxies,
which is roughly 10\% of the total sample, and they are located at
$z \lesssim 0.05$, corresponding to the depth $r \lesssim 150
\text{Mpc/h}$. We further assume that the measurement error for
line-of-sight velocity is around $20\%$, which is a typical error
for the fundamental plane distance measurement. We divide these
data into different redshift bins, and in each shell ($r$,
$r+dr$), we calculate the Fisher Matrix value for the CMN $F_{MM}$
from Eq.(\ref{like1}) which leads to the forecasted error of the
CMN as
\begin{equation}
\sigma_{M(r)} \simeq \frac{M(r)}{\sqrt{2N(r)}}\left[ 1+\frac{\sigma _{n}^{2}}{%
(u(r)/M(r))^{2}}\right],  \label{errorM}
\end{equation}
where $N(r)$ is the number of data points in the shell
($r$,$r+dr$), and $u(r)$ is the average bulk flow magnitude on
depth r. We plot these forecast data in the panel (a) of Fig.
\ref{6dFforecast}. Comparing with panel (b) of Fig.
\ref{datacompare}, We find the full range of the CMN data on
scales [10,150] (Mpc/h) from 6dF can improve the constraint on the
variation of the scale-dependent growth factor significantly. We
summarize the experimental conditions for future experiment that
can sharpen the CMN test: (1) there should be considerably more
galaxy samples ($\gtrsim 10^4$) on scales $[10,150] \text{Mpc/h}$;
(2) the smaller the measurement error $\sigma_{n}$ is, and the
larger sky area it covers, the better it can reduce the overall
error of $M$. One should also notice that, when the CMN data is
used to constrain cosmology, the cosmic variance (CV) due to the
limited volume of the survey should be taken
into account. The cosmic variance $\sigma_{\rm CV}(r)$ can be estimated as
\begin{equation}
\left[\frac{\sigma _{\rm CV}(r)}{M(r)}\right] ^{2}=\frac{1}{4}\left\{ \left[\frac{%
\sigma _{A}(r)}{A(r)}\right] ^{2}+\left[ \frac{\sigma
_{B}(r)}{B(r)}\right] ^{2}\right\}
\end{equation}
where $A$ and $B$ stand for
$\langle V^2(r)\rangle$ and $\langle \sigma^2(r)\rangle$
respectively, as defined in Eq (\ref{msv}), and $\sigma_A(r)$
and $\sigma_B(r)$ can be calculated as,

\begin{eqnarray}
\sigma _{A}(r)&=&\frac{H_0^2}{2\pi^2\sqrt{2\pi }}\int\frac{1}{\sqrt{%
k^{2}\Delta kV}}P_{\theta\theta}(k)W(kr)dk,\\
\sigma _{B}(r)&=&\frac{H_0^2}{2\pi^2\sqrt{2\pi }}\int\frac{1}{\sqrt{%
k^{2}\Delta kV}}P_{\theta\theta}(k)[1-W(kr)]dk.
\end{eqnarray}
where $\Delta k$
and $V$ are the width of the $k$ bin and the volume of the survey
respectively, and $W$ is the same window function as in Eq
(\ref{msv}). We found that for the current velocity surveys of
SBF, ENEAR, SN and SFI++$_F$, CV is 21, 19, 5 and 10\% of the
measurement error. For the 6dF survey, CV is smaller than 5\% of
the statistical errors in all the bins.

\section{A sensitive test of growth of structure}


Since the CMN measures the shape of the peculiar velocity power
spectrum $P_{\theta\theta}$ by design, it is sensitive to any
distortion of $P_{\theta\theta}$. In the $\Lambda$CDM model, the
growth is scale-independent. However, in the modified gravity
models, and models with massive neutrinos, the growth is
generically scale-dependent, thus $P_{\theta\theta}$ for these
models is a distorted version comparing with the $\Lambda$CDM
model, making the CMN an excellent tool to distinguish these
models from $\Lambda$CDM. In the following of this section, we
make separate discussion on the alternative theory of gravity and
massive neutrino, and their prospective discrimination from CMN.


\subsection{$f(R)$ gravity model}
In a viable $f(R)$ model, where the Einstein-Hilbert action is
extended to be a general function of the Ricci scalar $R$, the
effective value of Newton's constant $G_{\rm eff}$ has both time
and scale dependence, namely, $G_{\text{eff}}=\mu(a,k) G$, where
$\mu(a,k)=(1+\frac{4}{3}\lambda^2k^2 a^{4})/(1+\lambda^2k^2
a^{4})$, and $G$ is the Newton's constant in general relativity
(GR) (\citet{Zhao09,Giannantonio10}). The only free parameter is
$\lambda^2$, which quantifies the Compton wavelength of the scalar
field $f_R\equiv df/dR$ and characterizes the scale-dependent
growth rate.  It is more convenient to re-define a dimensionless
$B_{0}$ which is $B_{0}=2H^2_{0}\lambda^{2}/c^2$, and $B_0=0$ for
GR (\citet{Giannantonio10}).

In panels (b) and (c) of Fig. \ref{6dFforecast}, we show the
relative difference in $P_{\theta\theta}$ and $f$ with respective
to that in the $\Lambda$CDM model for a $f(R)$ model with
$B_{0}=10^{-4}$ (dashed lines). The growth rate for $f(R)$ model
is enhanced on scales of $k \gtrsim 0.01$ h/Mpc, thus the
integration of $P_{\theta\theta}(k)$ cumulates the deviation in
$k-$space and exhibits the substantial difference of CMN between
$\Lambda$CDM and $f(R)$ model. Unfortunately, this substantial
difference with $\Lambda$CDM is not observable by the current CMN
data due to the large errors, as shown in panel (b) of Fig.
\ref{datacompare}. However, as a result shown in panel (a) of Fig.
\ref{6dFforecast}, a $0.01\%$ of $B_0$ in $f(R)$ model can produce
a $20\%$ suppression in the CMN, which is potentially observable
by 6dF \footnote{We compare the forecast of 6dF constraint with
the $B_{0}=10^{-4}, 10^{-3}$ f(R) model in Fig.~\ref{6dFforecast}.
This is because the constraints from CMB, Type Ia supernovae and
ISW effect have already set $B_{0}<0.1$ \cite{Giannantonio10}}.



\begin{figure*}
\centerline{
\includegraphics[scale=0.15]{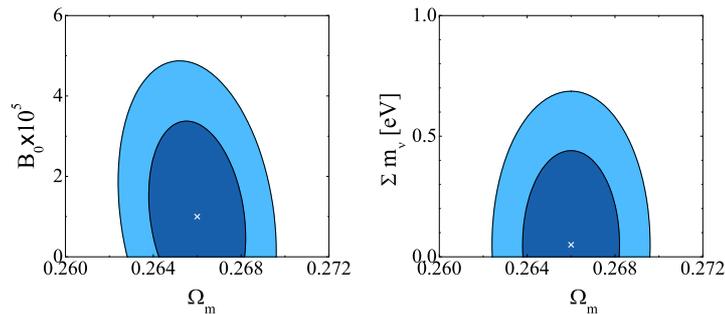}}
\caption{The 68 and 95\% CL contour plots for $B_0$, neutrino mass
and $\Omega_m$ derived from a combined dataset of WMAP, UNION2 and
6dF surveys using Fisher forecast. See the text for
details.}\label{fig:cont}
\end{figure*}

\subsection{Massive neutrinos}
Similarly, the CMN is sensitive to the neutrino mass. When
neutrinos became non-relativistic and the Universe was deeply in
the matter dominated era, neutrino thermal velocities damped out
the perturbation under the characteristic scale $k_{\text{nr}}
\simeq 0.018 \Omega^{1/2}_{m} (\sum m_{\nu}/1\text{eV})
\text{h/Mpc}$, suppressing the power spectrum on small scales. On
scales greater than $k_{\text{nr}}$, neutrinos affect the overall
expansion of the Universe and therefore shift the peak of power
spectrum to larger scales. In panels (b) and (c) of Fig.
\ref{6dFforecast}, we can see that neutrino with mass of $0.6
\text{eV}$ can suppress the power spectrum on scales of $k \gtrsim
0.01 \text{h/Mpc}$ significantly, which exactly falls in the
detection window of the CMN. Therefore, the cumulative `integral'
of $P_{\theta \theta}(k)$ in CMN can manifest the neutrino
free-streaming effect by enhancing its value on all scales.
Comparing with panel (b) of Fig. \ref{datacompare} and panel (a)
of Fig. \ref{6dFforecast}, we find that although the current data
is weak in discriminating the massive neutrino, future CMN data is
potentially able to distinguish the nonzero neutrino mass.

\subsection{Fisher Matrix forecast}
We are aware that the possible degeneracy among cosmological
parameters might dilute the constraints, therefore we employ a
Fisher matrix forecast (\citet{Tegmark:1996bz}) to study the
power of CMN on cosmological parameter constraints. The Fisher
matrix for CMN is,

\begin{equation}
F_{\mu\nu}=\sum_{i,j}\frac{\partial M(r_i)}{\partial p_{\mu}}{\rm
Cov}^{-1}[M(r_i),M(r_j)]\frac{\partial M(r_j)}{\partial p_{\nu}},
\end{equation}
where $p$ denotes the cosmological parameter, and Cov is the data
covariance matrix for CMN. We include the cosmic variance of CMN
in our analysis. We make a forecast for two models: $A=\{B_0,{\rm
CP}\},~~B=\{\Sigma m_\nu,{\rm CP}\}$ where CP is a set of basic
cosmological parameters: CP$=\{\Omega_m,\sigma_8,H_0,n_s\}$. We
choose the best-fit model for WMAP7 as a fiducial model
(\citet{Komatsu11}), namely, CP$=\{0.266,0.801,71,0.93\}$, and
choose $B_0=10^{-5}$ and $\Sigma m_\nu=0.05$ eV as fiducial values
for models A and B respectively. We also marginalize over the
nuisance parameter $C$ in Eq~(\ref{mach_theo}), whose fiducial value was evaluated using
the $\Lambda$CDM model. We also combine the Fisher matrices for
the current data of CMB (WMAP \citet{Komatsu11}) and SNe (UNION2
\citet{Amanullah10}) to improve the constraint (see details for
CMB and SNe Fisher matrices in \citet{Pogosian:2005ez,Zhao09}).
The result is shown in Fig \ref{fig:cont}. As we can see, the
constraints on $B_0$ and neutrino mass are largely improved when
the CMN data from 6dF is combined. Comparing with the constraints
without CMN but only WMAP7 combined with UNION2 data,
\citet{Giannantonio10} gives the 95\% CL. upper limits on $B_0$
as $B_0<0.4$ and \citet{Komatsu11} gives $\Sigma m_\nu<1.3
\textrm{eV}$. The constraints can be tightened to
$B_0<5\times10^{-5}$ and $\Sigma m_\nu<0.65 \textrm{eV}$ when the
CMN data from 6dF is included.




\section{Conclusion} We provide a statistical tool to measure
the CMN, and further demonstrate that it is a sensitive probe of
the structure growth. By design, the CMN is immune to the
uncertainty in the overall amplitude of the density perturbation,
and to linear galaxy bias. Also it is highly sensitive to any
scale-dependent distortion of $P_{\theta\theta}$ since any small
distortion can be amplified via the integral effect. Therefore it
is an excellent tool to test alternative theories of gravity, and
models with non-zero neutrino mass.

We first perform a likelihood analysis of the CMN from the current
peculiar velocity field data, and further confront the WMAP7
$\Lambda$CDM model and 1990s popular CDM model with the CMN data
from current and old observations. We confirm that the
$\Lambda$CDM model with WMAP 7-yr parameters is consistent with
current CMN data at $3\sigma$ CL. level. Based on our forecast for
6dF, we find that the CMN can improve the constraints on the
modified gravity parameter $B_0$ by 4 orders of magnitude for the fiducial $B_0$ value of $10^{-5}$, and it
can also tighten the present constraints on the neutrino mass. The
CMN information from future surveys, such as the \citet{ska},
will be more powerful to constrain cosmologies, especially for
modified gravity models, and models with massive neutrinos.

\acknowledgments
We would like to thank A.
Challinor, G. Efstathiou, K. Koyama and K. Masters for helpful
discussions. GBZ is supported by STFC grant ST/H002774/1.

\end{document}